\documentclass{article}

\makeatletter
\def\input@path{{../}{./}}
\makeatother

\usepackage[dblblindworkshop, final]{neurips_2026}

\usepackage[utf8]{inputenc}
\usepackage[T1]{fontenc}
\usepackage{hyperref}
\usepackage{url}
\usepackage{booktabs}
\usepackage{amsfonts}
\usepackage{nicefrac}
\usepackage{microtype}
\usepackage{xcolor}
\usepackage{graphicx}
\graphicspath{{./}{./figs/}{../}{../../experiments/fig_playground_outputs/}}
\usepackage{threeparttable}

\title{Proxy Metrics Can Hide Downstream Collapse:\\
A Case Study on Compressed Cellular Segmentation}

\workshoptitle{I Can't Believe It's Not Better (ICBINB)}

\author{%
  Sebasti\'an A.\ Cruz Romero\thanks{Corresponding author.} \\
  Capic\'u Technologies \\
  \texttt{scruzromero@capicu.ai} \\
}

\begin{document}

\maketitle

\begin{abstract}
Post-training compression of vision models is commonly screened using output-level proxies such as normalized RMSE (NRMSE), cosine similarity, and flow angular error. We evaluated post-training quantization of Cellpose--SAM, a promptless Segment--Anything model for cellular instance segmentation, on a stratified 245-field panel spanning BBBC038, BBBC039, and NIST iPSC images across density regimes. These proxies failed to reliably predict downstream mask retention under activation quantization. Angular error saturated at $90^\circ$ from A8 through A2, while cosine and NRMSE preserved rank-order information without providing an interpretable acceptance threshold. At W8A8-cal-QDQ, cosine similarity was $0.944$ despite $89/245$ empty masks; moreover, performance varied sharply by modality, with BBBC039 yielding $0/47$ empty masks versus $63/99$ for NIST iPSC fields at the same proxy footprint. Across five modality--density strata, cosine similarity and empty-mask rate showed weak rank agreement ($\rho \approx 0.3$), with the highest-cosine stratum producing the second-worst downstream outcome. Below A8, all modalities collapsed to $245/245$ empty fields. By contrast, ternary weight quantization produced a $12.08\times$ storage reduction and was correctly identified as catastrophic by the proxies. These results show that FP32-referenced output proxies are insufficient as standalone acceptance criteria for promptless dense-prediction models; compression should instead be validated against downstream retention stratified by deployment-relevant imaging conditions.
\end{abstract}

\section{Introduction}
\label{sec:intro}

Large vision models are increasingly being deployed under strict compute,
memory, and latency constraints, motivating post-training quantization (PTQ)
as a practical route to efficient inference. Across convolutional and
transformer architectures, PTQ workflows typically rely on numerical
proxies---including weight reconstruction error, activation discrepancy,
NRMSE, and cosine similarity relative to FP32 outputs---to screen candidate
compression schemes before downstream evaluation
\citep{jacob2018quantization,krishnamoorthi2018quantizing,nagel2020adaround,
li2021brecq,frantar2022gptq,dettmers2022llmint8,yao2022zeroquant,
lin2024awq,ma20241bit}. These proxies are attractive because they are
efficient, architecture-agnostic, and independent of task-specific
evaluation pipelines. Their implicit assumption is that numerical agreement
with FP32 provides a reliable surrogate for downstream model behavior.

For dense prediction systems, however, this assumption remains largely
untested. Segmentation foundation models used in scientific imaging do not
only require accurate intermediate representations; they require those
representations to remain compatible with downstream decoding procedures.
Small numerical deviations may therefore be benign, while changes that alter
decoder behavior can produce categorical failures despite apparently
acceptable proxy values. This distinction is particularly important for
cellular imaging workflows, where compressed models must operate across
heterogeneous modalities and biological regimes.

Here, we evaluate on Cellpose--SAM
\citep{stringer2025cellposesam}, a promptless segmentation foundation model,
under a deployment-oriented quantization protocol spanning 245 microscopy
fields from BBBC038, BBBC039, and NIST iPSC datasets. We find that standard
output-level proxies do not provide reliable deployment criteria for this
setting. Under whole-graph activation quantization, flow angular error
saturates at the same failure regime across activation widths, while cosine
similarity and NRMSE retain numerical variation without yielding an
interpretable acceptance threshold. More importantly, identical proxy
footprints can produce sharply different downstream outcomes across imaging
modalities: configurations that preserve segmentation quality in one
biological context can collapse in another. These results indicate that
compression safety for dense-prediction foundation models must be evaluated
through downstream retention criteria stratified by deployment conditions,
rather than by FP32-referenced similarity metrics alone.

\section{Related Work}
\label{sec:related}

\paragraph{Post-training quantization and its proxy metrics.} Modern PTQ recipes screen candidate schemes with cheap output-tensor proxies. \citet{jacob2018quantization} formalized integer-arithmetic quantized inference and introduced the reconstruction-error framing that has organized subsequent PTQ; \citet{krishnamoorthi2018quantizing} and \citet{nagel2021white} extended this into per-layer and per-tensor evaluation practices. Weight-side refinements including AdaRound~\citep{nagel2020adaround} and BRECQ~\citep{li2021brecq} minimize block-level reconstruction; on the transformer side, GPTQ~\citep{frantar2022gptq}, LLM.int8()~\citep{dettmers2022llmint8}, ZeroQuant~\citep{yao2022zeroquant}, AWQ~\citep{lin2024awq}, and the more recent 1.58-bit regime~\citep{ma20241bit} all rank candidates by activation- or weight-side reconstruction proxies. Sub-bit weight quantization studies~\citep{rastegari2016xnor,hubara2016binarized} follow the same convention. \citet{gholami2022survey} surveys the field, and the shared assumption throughout is that small numerical distortion at the tensor level implies small downstream distortion at the task level. Our case study addresses the direction in which that assumption fails.

\paragraph{Cellpose and Segment--Anything foundation models.} Cellpose introduced dynamics-based flow integration as the decoder for cellular instance segmentation~\citep{stringer2021cellpose,pachitariu2022cellpose2}, and Cellpose--SAM~\citep{stringer2025cellposesam} combines that decoder with the Segment--Anything visual encoder~\citep{kirillov2023sam,ravi2024sam2}. The promptless flow-integration decoding path is what places Cellpose--SAM in the same design regime as other decoder-heavy biomedical segmentation foundation models entering laboratory workflows~\citep{ma2024medsam}, and it is also what makes it sensitive to a class of numerical failure that mask-token or end-to-end differentiable decoders do not necessarily exhibit~\citep{weigert2020stardist}. Broader cellular deep learning has documented the growing role of segmentation foundation models in imaging pipelines~\citep{moen2019deep,meijering2020bird}.

\paragraph{Evaluation methodology and metric-validation guidance.} The medical-image-analysis community has, over the past several years, produced explicit guidance on the pitfalls of scalar per-pixel proxies and pooled reporting. \citet{maierhein2018rankings} showed that biomedical image-analysis competition rankings are unstable when hidden metric choices are varied; \citet{maierhein2022metrics} formalize per-modality and per-task metric recommendations under the Metrics Reloaded framework; and \citet{reinke2024pitfalls} enumerates specific proxy-vs-downstream failure modes. Segmentation-metric sample-size and confidence-interval heuristics~\citep{islam2018clt,varoquaux2022evaluating} and cluster-bootstrap inference~\citep{efron1994bootstrap,diciccio1996bootstrap} inform the retention criterion we use here. Pre-registration norms~\citep{nosek2018preregistration} motivate fixing the retention margin before any hold-out data are examined (in our study the margin is pre-specified internally rather than externally registered), and reproducibility reporting norms~\citep{pineau2021reproducibility} inform the release path for audit artifacts that make the retention verdicts re-derivable on independent data.

\section{Methods}
\label{sec:methods}

\paragraph{Model.} The compressed model in this study is Cellpose--SAM (checkpoint \texttt{cpsam\_v2}, $\sim$$305$\,M parameters, FP32 storage $1{,}162.07$ MiB), a promptless variant of the Segment--Anything family that emits per-pixel $dY$, $dX$, and cell-probability channels. Instance masks are produced by the dynamics-based seeding cascade introduced with Cellpose: local integration of the flow field yields attractor basins that are thresholded and reconciled into instance labels.

\paragraph{Panel.} Evaluation is performed on a stratified public-dataset panel: BBBC038 nuclei ($n=99$ fields spanning fluorescence, brightfield, and phase contrast~\citep{caicedo2019nucleus,ljosa2012bbbc}), BBBC039 U2OS fluorescence ($n=47$~\citep{ljosa2012bbbc}), and NIST induced-pluripotent-stem-cell tiles stratified over low, medium, and high culture-density regimes ($n=33$ per regime, $99$ total, aggregated over $3$ source-image experimental units~\citep{bajcsy2018enabling}). Whole-graph evaluations use all $245$ hold-out fields per scheme; mask outcomes are reported against both the FP32 counterpart (proxy-vs-outcome discordance) and supplied ground truth. Cluster-bootstrap catastrophic-failure intervals for the weight-only and mixed schemes are computed on a $176$-field subset ($107$ experimental units; BBBC038 $n=70$, BBBC039 $n=34$, NIST iPSC $n=72$) whose per-modality sizes were pre-specified to target the segmentation-metric sample-size design of~\citet{islam2018clt,varoquaux2022evaluating}. The subset is smaller than the whole-graph evaluation panel because proxy diagnostics can consume the full panel cheaply while cluster-bootstrap on paired retention requires the more expensive full-mask decode.

\paragraph{Retention criterion.} For each (scheme, dataset) pair we compute the mean change $\Delta$ in a downstream mask-level metric (paired instance F1 at IoU $0.5$; AP at IoU $\{0.5, \ldots, 0.95\}$) from FP32~\citep{lin2014coco} and form a $2{,}000$-draw cluster-bootstrap 95\% interval for $\mathbb{E}[\Delta]$ over experimental units. A scheme meets the criterion for that dataset if the entire interval lies above the pre-specified $-0.02$ margin. The margin was fixed before any hold-out evaluation. In addition, we report a per-scheme catastrophic-failure rate, the fraction of evaluation fields on which the compressed model's mask output falls below an absolute quality floor, with an accompanying bootstrap 95\% interval over experimental units.

\paragraph{Proxies.} We study three complementary output-tensor proxies used across the PTQ literature. Per-field NRMSE is the L2 error against the FP32 reference on the concatenated $dY$, $dX$, cell-probability output normalized by FP32 L2 norm~\citep{jacob2018quantization,nagel2021white,gholami2022survey}. Per-field cosine similarity is the cosine between the flattened compressed and FP32 outputs; and per-field median flow angular error is a decoder-aware proxy on the direction of the decoded flow field, computed as the median per-pixel angle between the compressed and FP32 flow vectors.

\paragraph{Compression schemes.} We evaluate three weight-only schemes (\mbox{W8A16}, \mbox{W4A16-G64} with group size $64$, and ternary \mbox{W2A16-G64}); a calibrated whole-graph \mbox{W8A8-cal-QDQ} together with a whole-graph integer \mbox{W8A8}; and four low-bit-width whole-graph configurations (\mbox{W8A4}, \mbox{W8A4} with softmax-A8 stabilization, \mbox{W8A2}, and \mbox{W8A2} with softmax-A8). A mixed \mbox{W4/W8} scheme retains at INT8 the four most sensitive operators identified by a one-operator-at-a-time W4 perturbation over $100$ eligible Linear/Conv2d operators on the development split. W8A8-cal-QDQ was calibrated using five separate microscopy fields: BBBC038 nuclei, BBBC039 U2OS nuclei, and NIST iPSCs at low, medium, and high cell densities.

\begin{figure*}[t]
\centering
\includegraphics[width=\textwidth]{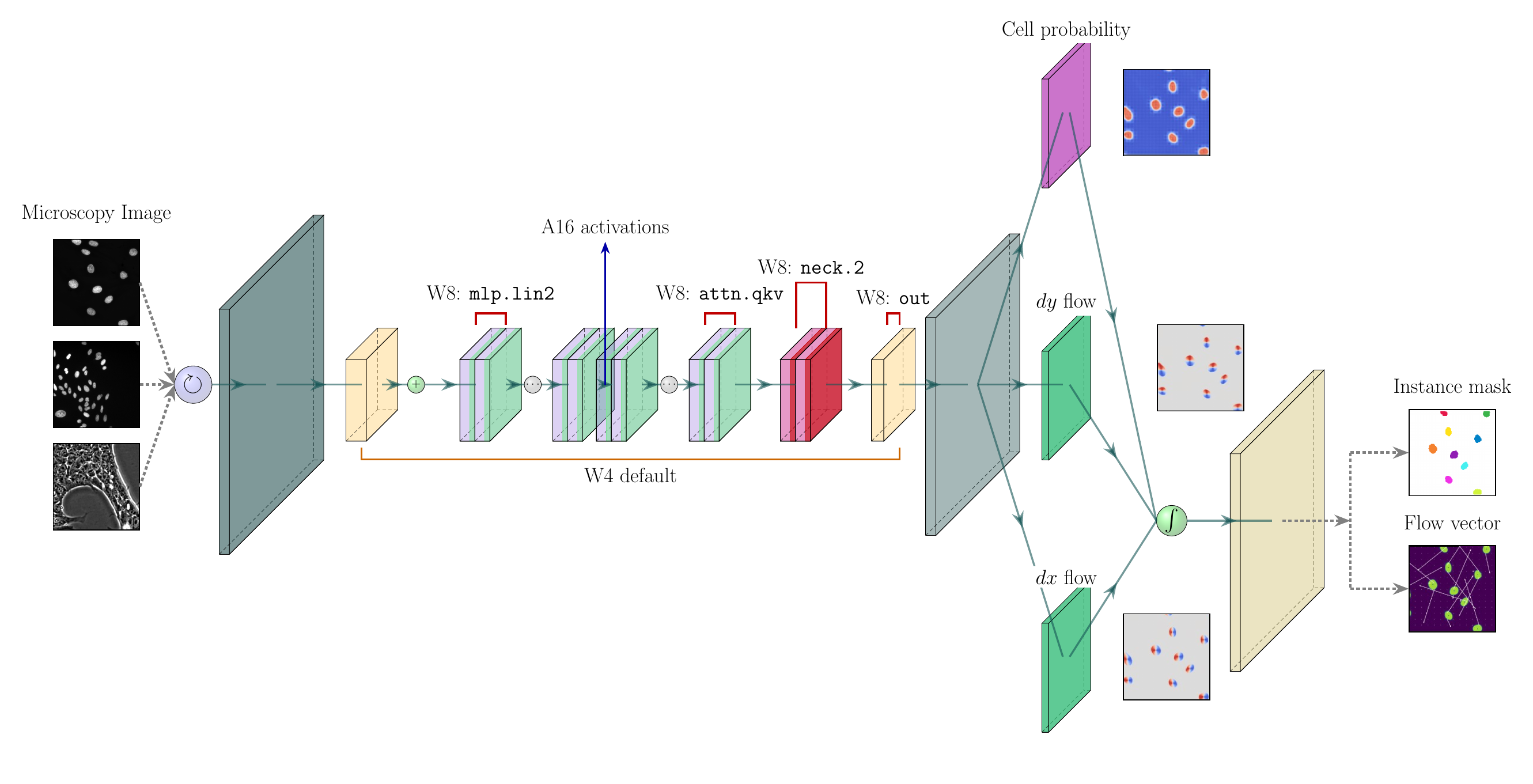}
\caption{\textbf{Where the proxies live inside Cellpose--SAM.} A microscopy field (fluorescence, brightfield, or phase contrast)
  passes through the SAM visual encoder into the flow-integration decoder. The default weight schedule is W4 (orange bracket); illustrating as an example, the
  sensitivity-guided mixed scheme retains four operator classes at W8 (\texttt{mlp.lin2}, \texttt{attn.qkv}, \texttt{neck.2}, and
  \texttt{out}) and activations are held at A16. The decoder emits a cell-probability map and a $(dY, dX)$ flow field; flow
  integration $\int$ produces the decoded instance mask. Output-tensor proxies (NRMSE, cosine similarity, median flow angular
  error) are computed on the raw pre-integration tensors marked here, i.e., before the seed-detection step of the
  flow-integration cascade where whole-graph W8A8 collapse localizes.}
\label{fig:architecture-closeup}
\end{figure*}

\section{Results}
\label{sec:results}

\begin{figure*}[t]
\centering
\includegraphics[width=\textwidth]{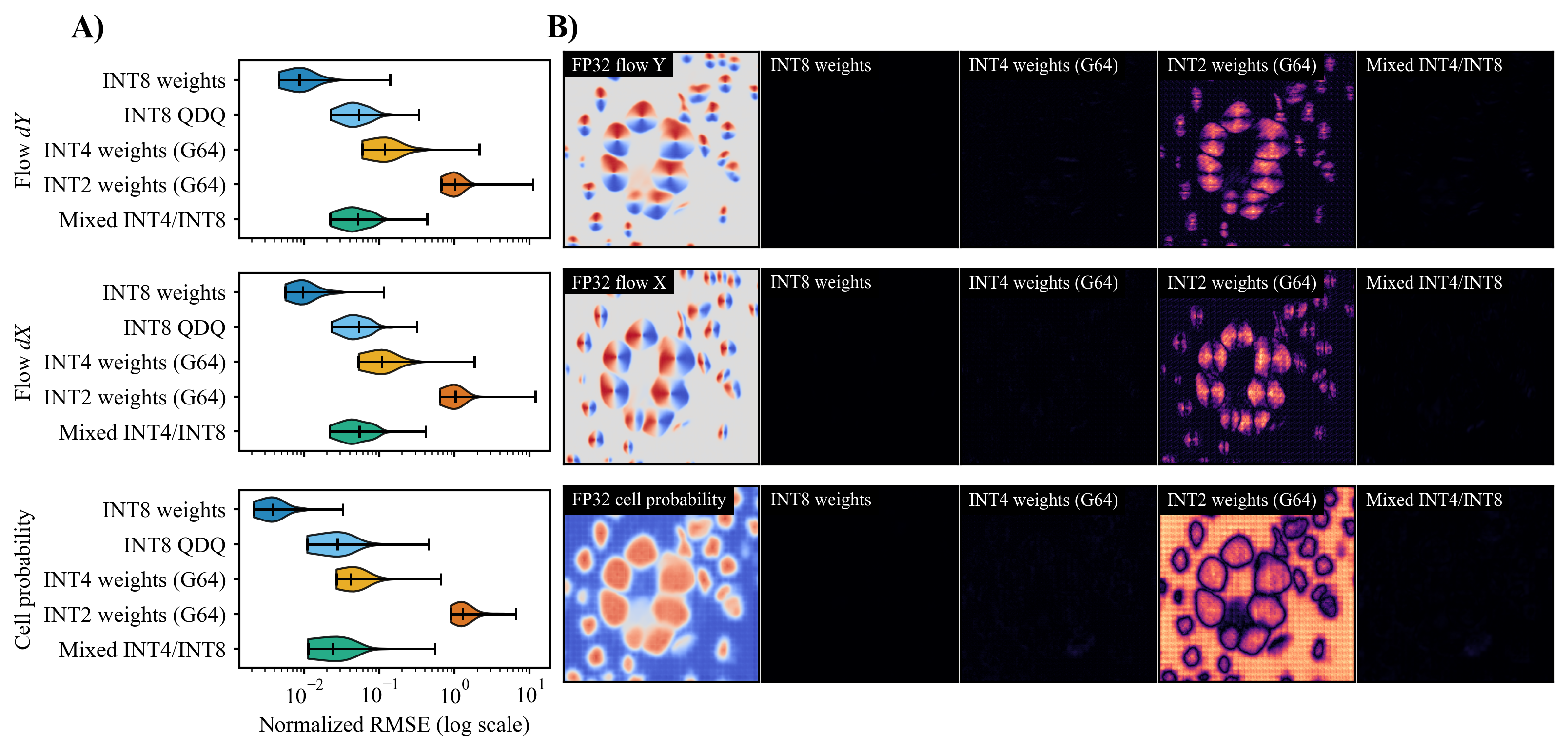}
\caption{\textbf{Raw-output proxies on the weight-only path.} \textbf{A)} per-field raw-output normalized RMSE against FP32
  (log scale) split by output channel ($dY$ flow, $dX$ flow, cell probability) for INT8 weights (W8A16), INT8-QDQ,
  INT4 weights with group size $64$ (W4A16-G64), INT2 weights with group size $64$ (W2A16-G64), and the sensitivity-guided
  mixed INT4/INT8 scheme. \textbf{B)} matched qualitative composite of FP32 flow-$Y$, flow-$X$, and cell-probability maps
  against the four quantized configurations on a representative field. NRMSE separates ternary weights from the other
  schemes but does not resolve the mixed INT4/INT8 candidate from INT4, motivating the downstream retention check reported
  in Table~\ref{tab:weight-only}.}
\label{fig:raw-output-proxies}
\end{figure*}

Table~\ref{tab:weight-only} and Figure~\ref{fig:raw-output-proxies} report the weight-only bit-width sweep. On this path the three proxies and the downstream retention outcome are broadly monotone with bit width: median NRMSE rises from $0.008$ at W8A16 to $0.095$ at W4A16-G64 to $1.038$ at W2A16-G64, cosine similarity falls from $1.000$ to $0.997$ to $0.423$, and the bootstrap 95\% catastrophic-failure interval widens from $[0.00, 0.00]$ to $[0.00, 0.001]$ to $[0.95, 1.00]$. On this path a standard PTQ workflow is defensible: a proxy screen at $\mathrm{NRMSE}\!\sim\!1$ would reject W2A16-G64 for the same reason the retention criterion does, and the $12.08\times$ storage reduction that ternary weights promise does not override the catastrophic-failure rate on this task, bootstrap point estimate $97.9\%$ over experimental units with raw fraction $169/176 = 96.0\%$.

\begin{table}[t]
\centering
\small
\caption{Weight-only compression evaluation on the 176-field retention
subset (107 experimental units). Compression statistics, reconstruction
proxies, and catastrophic-failure rates are reported for each quantization
scheme. Catastrophic rates use 2,000-draw cluster-bootstrap percentile
intervals over experimental units.}
\label{tab:weight-only}
\begin{tabular}{lccccc}
\toprule
&
\multicolumn{2}{c}{Compression} &
\multicolumn{3}{c}{Reconstruction quality} \\
\cmidrule(lr){2-3}
\cmidrule(lr){4-6}
Scheme &
Size (MiB) &
Ratio &
NRMSE &
Range &
Catastrophic rate [95\% CI] \\
\midrule
FP32 (reference) & 1162.07 & 1.00$\times$ & --- & --- & --- \\
W8A16            & 295.78  & 3.93$\times$ & 0.008 & 0.002--0.059 & 0.000 [0.000, 0.028]$^{\dagger}$ \\
W4A16-G64        & 168.47  & 6.90$\times$ & 0.095 & 0.027--0.722 & 0.001 [0.000, 0.001] \\
W2A16-G64        & 96.21   & 12.08$\times$ & 1.038 & 0.680--4.696 & 0.979 [0.949, 1.000] \\
\bottomrule
\end{tabular}

\vspace{2pt}
\begin{minipage}{0.94\linewidth}
\footnotesize
$^{\dagger}$ Zero observed failures; percentile bootstrap is replaced by the
rule-of-three upper bound $3/107$ over experimental units.
\end{minipage}
\end{table}

\begin{figure*}[t]
\centering
\includegraphics[width=.78\textwidth]{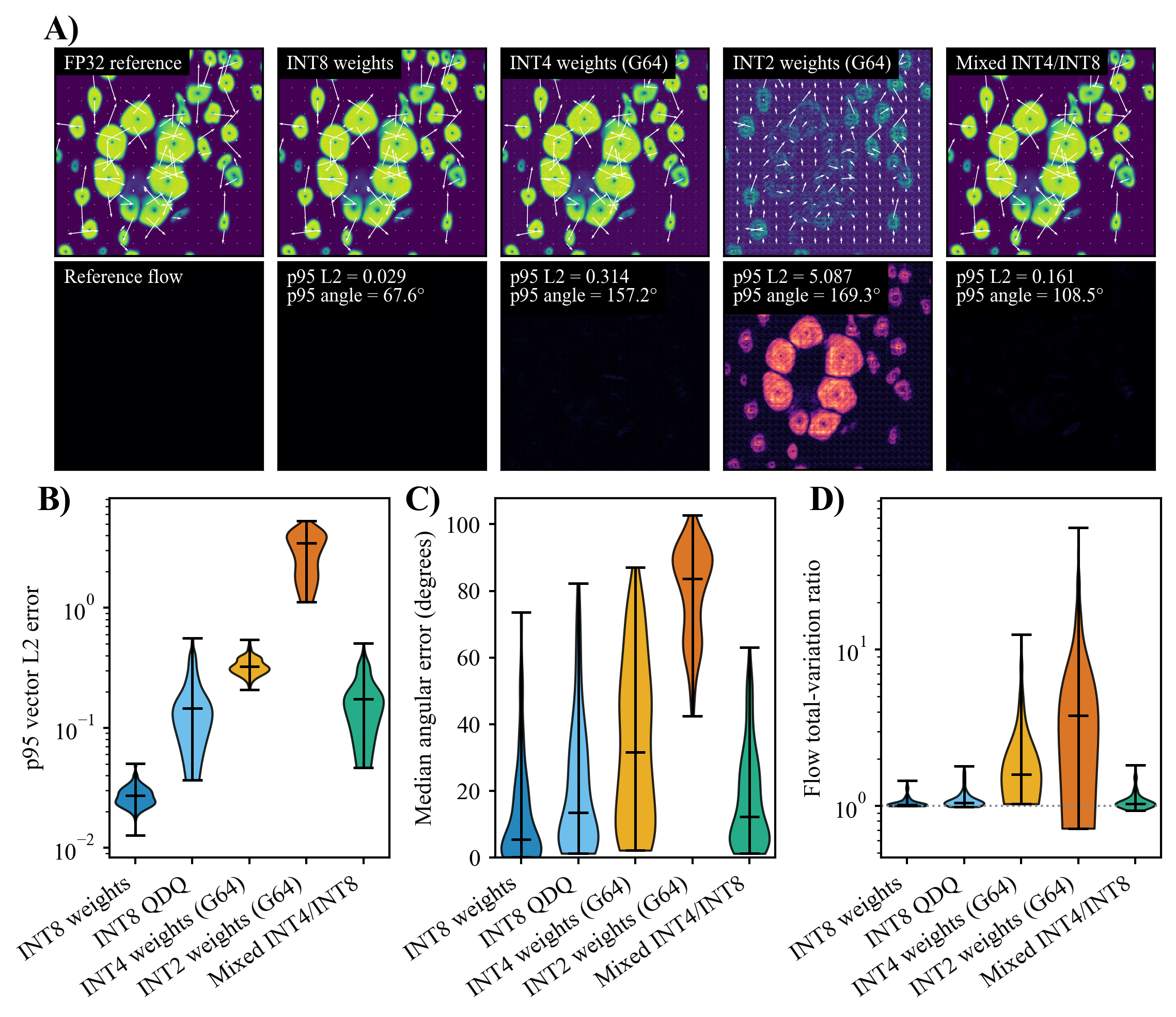}
\caption{\textbf{Decoder-aware flow-field proxies.}
\textbf{A}, qualitative decoded flow fields for FP32 and four quantized
configurations, with per-field 95th-percentile L2 error, angular error,
and zero-flow plateau fraction.
\textbf{B--D}, retention-panel distributions of
\textbf{B} 95th-percentile vector L2 error,
\textbf{C} median angular error, and
\textbf{D} decoded-flow total-variation ratio.
W2A16-G64 shifts angular error toward $90^\circ$, consistent with
a categorical flow-field collapse that is not captured by scalar
output-tensor similarity metrics.}
\label{fig:flow-proxies}
\end{figure*}

\begin{figure*}[t]
\centering
\includegraphics[width=.85\textwidth]{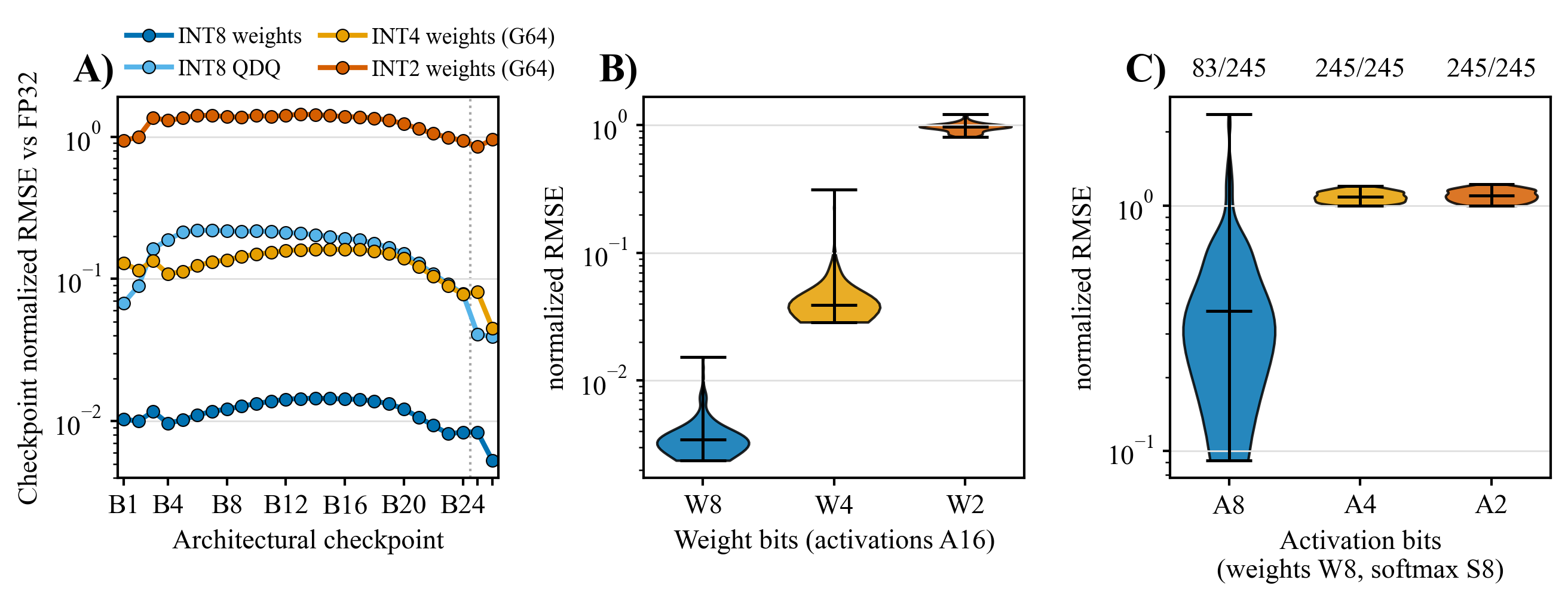}
\caption{\textbf{Where in the graph the quantization budget spends.} \textbf{A)} checkpoint-level normalized RMSE against
  FP32 through the $26$ architectural checkpoints (B1--B26, comprising the ViT-L encoder's $24$ transformer blocks together with the patch-embedding and neck taps), stratified by scheme; the calibrated W8A8-cal-QDQ curve tracks the W8
  weight-only curve for most of the encoder but diverges near the neck. \textbf{B)} per-field raw-output NRMSE aggregated
  by weight bit width at fixed A16 activations, showing the weight-only sensitivity floor. \textbf{C)} activation-bit sweep
  at fixed W8 (softmax stabilization pinned to A8): $89/245$ hold-out fields produce empty masks at A8 (W8A8-cal-QDQ, and
  $83/245$ at W8A8-full-int) and $245/245$ produce empty masks at both A4 and A2, so the proxy in panel B, only
  $\sim$$2\times$ larger between A4 and A8, with median flow angular error already at $90^\circ$ at A8, does not track
  the downstream shift from partial to categorical collapse reported in Tables~\ref{tab:whole-graph}
  and~\ref{tab:whole-graph-biology}.}
\label{fig:whole-graph-collapse}
\end{figure*}

Table~\ref{tab:whole-graph} together with Figures~\ref{fig:flow-proxies} and~\ref{fig:whole-graph-collapse} reports the whole-graph path, and this is where the three proxies fail in complementary ways. Angular error is a resolution failure: median flow angular error is $\sim$$90^\circ$ (uniform-random against FP32) at W8A8-cal-QDQ, W8A8-full-int, W8A4, and W8A2 alike, so at the pooled panel level the proxy cannot separate partial from categorical downstream collapse. Cosine and NRMSE are a threshold failure: cosine cleanly separates A8 ($0.944$) from A4 ($\sim$$0.2$) and A2 ($\sim$$0.0$), and NRMSE moves $\sim$$2\times$ from $0.496$ at W8A8-cal-QDQ to $1.100$ at W8A2, so both carry rank-order information, but a cosine of $0.944$ reads as excellent to a practitioner while coexisting with $89/245$ empty masks, and neither has a defensible accept threshold near the operating regime. Downstream, at the W8A8-cal-QDQ proxy footprint the model emits zero instances on $89/245$ cases with median AP$_{50}$ (against FP32) $= 0.100$, a partial collapse. At W8A4 and W8A2 emission drops to $0$ instances on $245/245$ cases with median AP$_{50}$ $= 0.000$, so at essentially the same angular-error proxy value the outcome shifts from partial to categorical. Also visible in Table~\ref{tab:whole-graph} but unexplained by the calibration construction: W8A8-full-int has a median AP$_{50}$ of $0.521$ against $0.100$ for the calibrated W8A8-cal-QDQ control on the same $245$ fields, a $\sim$$5\times$ reversal opposite to what the harsher rounding path of full-integer arithmetic on top of QDQ would predict. We report the ordering but do not resolve it in this study.

The discordance is biology-conditioned, and stratification recovers signal the pooled row of Table~\ref{tab:whole-graph} hides. Table~\ref{tab:whole-graph-biology} stratifies the W8A8-cal-QDQ result by (dataset, regime); at the stratum level median angular error is $58^\circ$ on BBBC039 and $87^\circ$ on NIST high density against $90^\circ$ on the other three strata, so the panel-level $90^\circ$ figure is a pooling artifact and the proxy is not saturated everywhere. Even so, at the stratum level the proxy is non-monotone in the outcome. Sorted by cosine (best proxy first), the five strata rank as NIST low ($0.972$, $24/33 = 72.7\%$ empty), BBBC039 ($0.963$, $0/47 = 0\%$), NIST medium ($0.956$, $11/33 = 33.3\%$), BBBC038 ($0.942$, $26/99 = 26.3\%$), NIST high ($0.421$, $28/33 = 84.8\%$); Spearman rank correlation between cosine and empty-mask rate across the five strata is $\rho \approx 0.3$, near-random agreement, and the best-cosine stratum in the panel produces the second-worst downstream outcome. BBBC039 U2OS fluorescence retains full instance quality ($0/47$ empty, median AP$_{50}$ $= 0.927$) and is also the one stratum where the angular-error proxy has not saturated. NIST iPSC high density (cosine $0.421$) is the only stratum where cosine registers damage, and it collapses regardless. The axis on which the proxies mislead is imaging modality $\times$ density regime, precisely the axis a compressed encoder is deployed against.

\begin{table}[t]
\centering
\small
\caption{Whole-graph activation quantization evaluation on 245 hold-out
fields. Activation agreement is measured against the FP32 reference using
NRMSE, cosine similarity, and flow angular error; downstream performance is
reported by empty-mask frequency and instance detection quality.}
\label{tab:whole-graph}
\begin{tabular}{lcccccc}
\toprule
& \multicolumn{4}{c}{Activation agreement} &
\multicolumn{2}{c}{Detection outcome} \\
\cmidrule(lr){2-5}
\cmidrule(lr){6-7}
Scheme &
NRMSE &
Range &
Cosine &
Flow err. &
Empty masks &
AP$_{50}$ \\
\midrule
W8A8-cal-QDQ     & 0.496 & 0.11--2.69 & 0.944 & 90$^\circ$ & 89/245  & 0.100 \\
W8A8-full-int    & 0.372 & 0.09--2.36 & 0.963 & 90$^\circ$ & 83/245  & 0.521 \\
W8A4-full-int-S8 & 1.088 & 1.00--1.20 & 0.219 & 90$^\circ$ & 245/245 & 0.000 \\
W8A4-full-int    & 1.088 & 1.00--1.20 & 0.204 & 90$^\circ$ & 245/245 & 0.000 \\
W8A2-full-int-S8 & 1.100 & 1.00--1.22 & 0.000 & 90$^\circ$ & 245/245 & 0.000 \\
W8A2-full-int    & 1.100 & 1.00--1.22 & 0.000 & 90$^\circ$ & 245/245 & 0.000 \\
\bottomrule
\end{tabular}
\end{table}

\begin{table}[t]
\centering
\small
\caption{Biology-conditioned proxy discordance for W8A8-cal-QDQ on the
245-field hold-out panel relative to FP32. Rows are grouped by imaging
modality and density regime. Metrics include activation agreement proxies
(NRMSE, cosine similarity, and flow angular error) and downstream mask
retention.}
\label{tab:whole-graph-biology}
\begin{tabular}{llccccc}
\toprule
Modality & Regime & $n$ & NRMSE & Cosine & Flow err. & Empty (AP$_{50}$ med) \\
\midrule
BBBC039   & U2OS fluorescence     & 47 & 0.482 & 0.963 & 58$^\circ$ & 0/47  \phantom{0} (0.927) \\
BBBC038   & Heterogeneous nuclei  & 99 & 0.563 & 0.942 & 90$^\circ$ & 26/99 \phantom{0} (0.536) \\
NIST iPSC & Medium density        & 33 & 0.381 & 0.956 & 90$^\circ$ & 11/33 \phantom{0} (0.000) \\
NIST iPSC & Low density           & 33 & 0.420 & 0.972 & 90$^\circ$ & 24/33 \phantom{0} (0.000) \\
NIST iPSC & High density          & 33 & 0.927 & 0.421 & 87$^\circ$ & 28/33 \phantom{0} (0.000) \\
\bottomrule
\end{tabular}
\end{table}

A widespread PTQ heuristic identifies the most-sensitive operators by a proxy criterion and preserves their precision, and Table~\ref{tab:mixed-precision} shows where this heuristic still works on our task. A one-operator-at-a-time W4 perturbation over $100$ eligible Linear/Conv2d operators isolates four highest-sensitivity operators (\texttt{encoder.blocks.0.mlp.lin2}, \texttt{encoder.blocks.23.attn.qkv}, \texttt{encoder.neck.2}, \texttt{out}); retaining these at INT8 while the remainder use W4 gives a $6.76\times$ weight-storage reduction ($171.86$ MiB), and delivers $0/176$ observed catastrophic fields (rule-of-three upper bound $3/107 \approx 0.028$ over experimental units), the same zero-event outcome as W8A16 at this sample size. The proxy-driven selection therefore transfers within the weight-only regime. Applying the same operator selection under a whole-graph activation-quantization envelope does not recover the collapsed configurations of Table~\ref{tab:whole-graph}: the retained INT8 exceptions do not fix a failure whose mechanism is, we hypothesize, post-decoder seed detection rather than weight reconstruction at any individual operator.

\begin{table}[t]
\centering
\small
\begin{threeparttable}
\caption{Sensitivity-guided mixed \mbox{W4/W8} scheme. Four INT8 exceptions are selected from $100$ candidate operators by a one-operator-at-a-time W4 perturbation on the development split; the remainder use W4. The scheme achieves a $6.76\times$ weight-storage reduction and $0/176$ observed catastrophic fields (rule-of-three upper bound $0.028$ over $107$ experimental units), the same zero-event outcome as W8A16 at this sample size.}
\label{tab:mixed-precision}
\begin{tabular}{lcccc}
\toprule
&
\multicolumn{2}{c}{Compression} &
\multicolumn{2}{c}{Robustness} \\
\cmidrule(lr){2-3}
\cmidrule(lr){4-5}
Scheme &
Size (MiB) &
Ratio &
Catastrophic rate [95\% CI] &
INT8 exceptions \\
\midrule
mixed-W4-sensitive-W8 &
171.86 &
6.76$\times$ &
0.000 [0.000, 0.028]$^{\dagger}$ &
4 layers\tnote{a} \\
\bottomrule
\end{tabular}

\begin{tablenotes}[flushleft]
\footnotesize
\item[$^{\dagger}$] Rule-of-three upper bound $3/107$ over experimental units, in place of the degenerate zero-event bootstrap interval.
\item[a] INT8 exceptions:
\texttt{blocks.0.mlp.lin2},
\texttt{blocks.23.attn.qkv},
\texttt{neck.2}, and
\texttt{out}.
\end{tablenotes}

\end{threeparttable}
\end{table}

\section{Discussion \& Conclusion}
\label{sec:discussion}

The three proxies studied (NRMSE, cosine similarity, and median flow angular error) fail in complementary ways under whole-graph activation quantization. Angular error is a resolution failure: median flow angular error is $90^\circ$ at W8A8-cal-QDQ, W8A8-full-int, W8A4, and W8A2 alike; a decoded flow field that is uniform-random against FP32 is the boundary the angular-error proxy can measure, and every whole-graph configuration we test is already at that boundary at the pooled panel level, so the proxy cannot separate partial ($89/245$ empty at W8A8) from categorical ($245/245$ at sub-A8) collapse. Stratification recovers signal the pooled panel hides: median angular error is $58^\circ$ on BBBC039 and $87^\circ$ on NIST high density against $90^\circ$ on the other three strata (Table~\ref{tab:whole-graph-biology}). Cosine and NRMSE are a threshold failure: cosine remains $\geq 0.94$ at W8A8-cal-QDQ, compatible on our panel with $89/245$ empty-mask cases and a median AP$_{50}$ of only $0.100$; NRMSE moves $\sim$$2\times$ across activation widths that separate partial from categorical downstream collapse. Both carry rank information (cosine separates A8, A4, and A2 cleanly), but a cosine of $0.944$ reads as excellent to a practitioner while coexisting with $36\%$ empty masks, and neither proxy has a defensible accept threshold near the operating regime.

The most consequential piece of the picture is that the downstream discordance is biology-conditioned and non-monotone in the proxy. Sorted by cosine (best proxy first) the five per-modality strata in Table~\ref{tab:whole-graph-biology} rank as NIST low ($0.972$, $72.7\%$ empty), BBBC039 ($0.963$, $0\%$), NIST medium ($0.956$, $33.3\%$), BBBC038 ($0.942$, $26.3\%$), NIST high ($0.421$, $84.8\%$); Spearman rank correlation between cosine and empty-mask rate is $\rho \approx 0.3$, near-random agreement, and the highest-cosine stratum in the panel produces the second-worst downstream outcome. On BBBC039 U2OS fluorescence, the one stratum where the angular-error proxy has not saturated ($58^\circ$), the whole-graph W8A8-cal-QDQ scheme is essentially non-disruptive downstream and the proxies (NRMSE $0.482$, cosine $0.963$) match. NIST iPSC high-density (cosine $0.421$) is the only sub-stratum where cosine registers damage, and it collapses regardless. The axis on which the proxies mislead is the axis on which the compressed model is deployed. This is not a defect in any individual proxy; it is a structural consequence of comparing pixel- or channel-wise residuals against a bounded reference through a non-linear post-hoc decoding cascade whose invariances the proxies do not represent. We hypothesize the specific mechanism is that seed detection in the flow-integration cascade requires a directionally coherent flow field; our data are consistent with this account without isolating it, and a decisive test is a channel-substitution experiment that pairs FP32 flow with quantized cell-probability (and the inverse) to identify which channel's quantization is dispositive. We report this substitution experiment as an open follow-up.

For compression evaluation of promptless dense-prediction foundation models, we recommend pre-registering the downstream retention margin before hold-out evaluation, reporting per-(scheme, dataset, regime) verdicts with cluster-bootstrap intervals rather than pooled retention numbers, and releasing per-field metric tables so that verdicts can be re-derived on independent data. On our task, this protocol accepts weight-only \mbox{W8A16} and a sensitivity-guided mixed \mbox{W4/W8} scheme that achieves $6.76\times$ weight-storage reduction with $0/176$ observed catastrophic fields (rule-of-three upper bound $0.028$ over $107$ experimental units); it rejects ternary weight-only quantization at any storage advantage; and it rejects every whole-graph activation-quantization configuration tested despite proxy profiles that would tempt a practitioner to accept the W8A8 configurations, most sharply on the iPSC modality that motivates the deployment.

\section{Limitations}
\label{sec:limits}

The findings scope to promptless flow-integration decoders and to a mask-level AP retention criterion; pipelines with different decoder families or that consume the raw flow field directly should re-derive the mapping from proxy to retention rather than inherit it. The retention margin used here ($-0.02$ mean AP change from FP32) is calibrated to segmentation-driven monitoring workflows that tolerate a bounded AP drop; deployments with tighter numerical tolerances (regulatory release testing) require a tighter margin, and the protocol scales but the verdicts do not transfer.

\begin{ack}
Do {\bf not} include this section in the anonymized submission, only in the final paper. You can use the \texttt{ack} environment provided in the style file to automatically hide this section in the anonymized submission.
\end{ack}

\bibliographystyle{plainnat}
\bibliography{icbinb_wrksp_cruzromero/ref}

@inproceedings{kirillov2023sam,
  title     = {Segment Anything},
  author    = {Kirillov, Alexander and Mintun, Eric and Ravi, Nikhila and Mao, Hanzi and
               Rolland, Chloe and Gustafson, Laura and Xiao, Tete and Whitehead, Spencer and
               Berg, Alexander C. and Lo, Wan-Yen and Doll{\'a}r, Piotr and Girshick, Ross},
  booktitle = {IEEE/CVF International Conference on Computer Vision (ICCV)},
  year      = {2023}
}

@article{ravi2024sam2,
  title   = {{SAM 2}: Segment Anything in Images and Videos},
  author  = {Ravi, Nikhila and Gabeur, Valentin and Hu, Yuan-Ting and Hu, Ronghang and
             Ryali, Chaitanya and Ma, Tengyu and Khedr, Haitham and R{\"a}dle, Roman and
             Rolland, Chloe and Gustafson, Laura and others},
  journal = {arXiv preprint arXiv:2408.00714},
  year    = {2024}
}

@article{stringer2021cellpose,
  title   = {Cellpose: a generalist algorithm for cellular segmentation},
  author  = {Stringer, Carsen and Wang, Tim and Michaelos, Michalis and Pachitariu, Marius},
  journal = {Nature Methods},
  volume  = {18},
  number  = {1},
  pages   = {100--106},
  year    = {2021}
}

@article{pachitariu2022cellpose2,
  title   = {Cellpose 2.0: how to train your own model},
  author  = {Pachitariu, Marius and Stringer, Carsen},
  journal = {Nature Methods},
  volume  = {19},
  number  = {12},
  pages   = {1634--1641},
  year    = {2022}
}

@misc{stringer2025cellposesam,
  title  = {Cellpose--{SAM}: a foundation model for generalist cellular segmentation},
  author = {Stringer, Carsen and Pachitariu, Marius},
  year   = {2025},
  note   = {Preprint. Cellpose--SAM release notes and model checkpoint \texttt{cpsam\_v2}.}
}

@article{ma2024medsam,
  title   = {Segment anything in medical images},
  author  = {Ma, Jun and He, Yuting and Li, Feifei and Han, Lin and You, Chenyu and Wang, Bo},
  journal = {Nature Communications},
  volume  = {15},
  pages   = {654},
  year    = {2024}
}

@article{weigert2020stardist,
  title   = {Star-convex polyhedra for 3{D} object detection and segmentation in microscopy},
  author  = {Weigert, Martin and Schmidt, Uwe and Haase, Robert and Sugawara, Ko and
             Myers, Gene},
  journal = {IEEE/CVF Winter Conference on Applications of Computer Vision (WACV)},
  year    = {2020}
}

@article{moen2019deep,
  title   = {Deep learning for cellular image analysis},
  author  = {Moen, Erick and Bannon, Dylan and Kudo, Takamasa and Graf, William and
             Covert, Markus and {Van Valen}, David},
  journal = {Nature Methods},
  volume  = {16},
  number  = {12},
  pages   = {1233--1246},
  year    = {2019}
}

@article{meijering2020bird,
  title   = {A bird's-eye view of deep learning in bioimage analysis},
  author  = {Meijering, Erik},
  journal = {Computational and Structural Biotechnology Journal},
  volume  = {18},
  pages   = {2312--2325},
  year    = {2020}
}

@article{caicedo2019nucleus,
  title   = {Nucleus segmentation across imaging experiments: the 2018 {D}ata {S}cience {B}owl},
  author  = {Caicedo, Juan C. and Goodman, Allen and Karhohs, Kyle W. and Cimini, Beth A. and
             Ackerman, Jeanelle and Haghighi, Marzieh and Heng, CherKeng and Becker, Tim and
             Doan, Minh and McQuin, Claire and Rohban, Mohammad and Singh, Shantanu and
             Carpenter, Anne E.},
  journal = {Nature Methods},
  volume  = {16},
  number  = {12},
  pages   = {1247--1253},
  year    = {2019}
}

@article{ljosa2012bbbc,
  title   = {Annotated high-throughput microscopy image sets for validation ({BBBC})},
  author  = {Ljosa, Vebjorn and Sokolnicki, Katherine L. and Carpenter, Anne E.},
  journal = {Nature Methods},
  volume  = {9},
  number  = {7},
  pages   = {637},
  year    = {2012}
}

@article{bajcsy2018enabling,
  title   = {Enabling interactive characterization of stem-cell colony morphology using large microscopy images},
  author  = {Bajcsy, Peter and Vandecreme, Antoine and Amelot, Julien and Chalfoun, Joe and
             Majurski, Michael and Brady, Mary},
  journal = {Cell Reports Methods},
  year    = {2018},
  note    = {Referenced here for the NIST iPSC low/medium/high density regime tiles used in this work.}
}

@article{jacob2018quantization,
  title   = {Quantization and Training of Neural Networks for Efficient Integer-Arithmetic-Only Inference},
  author  = {Jacob, Benoit and Kligys, Skirmantas and Chen, Bo and Zhu, Menglong and Tang, Matthew and
             Howard, Andrew and Adam, Hartwig and Kalenichenko, Dmitry},
  journal = {IEEE/CVF Conference on Computer Vision and Pattern Recognition (CVPR)},
  year    = {2018}
}

@article{nagel2021white,
  title   = {A White Paper on Neural Network Quantization},
  author  = {Nagel, Markus and Fournarakis, Marios and Amjad, Rana Ali and Bondarenko, Yelysei and
             van Baalen, Mart and Blankevoort, Tijmen},
  journal = {arXiv preprint arXiv:2106.08295},
  year    = {2021}
}

@article{gholami2022survey,
  title   = {A survey of quantization methods for efficient neural network inference},
  author  = {Gholami, Amir and Kim, Sehoon and Dong, Zhen and Yao, Zhewei and
             Mahoney, Michael W. and Keutzer, Kurt},
  journal = {Low-Power Computer Vision, Chapman and Hall/CRC},
  year    = {2022}
}

@techreport{krishnamoorthi2018quantizing,
  title  = {Quantizing deep convolutional networks for efficient inference: A whitepaper},
  author = {Krishnamoorthi, Raghuraman},
  year   = {2018},
  institution = {arXiv preprint arXiv:1806.08342}
}

@inproceedings{nagel2020adaround,
  title     = {Up or down? {A}daptive rounding for post-training quantization},
  author    = {Nagel, Markus and Amjad, Rana Ali and van Baalen, Mart and Louizos, Christos and
               Blankevoort, Tijmen},
  booktitle = {International Conference on Machine Learning (ICML)},
  year      = {2020}
}

@inproceedings{li2021brecq,
  title     = {{BRECQ}: Pushing the limit of post-training quantization by block reconstruction},
  author    = {Li, Yuhang and Gong, Ruihao and Tan, Xu and Yang, Yang and Hu, Peng and
               Zhang, Qi and Yu, Fengwei and Wang, Wei and Gu, Shi},
  booktitle = {International Conference on Learning Representations (ICLR)},
  year      = {2021}
}

@inproceedings{frantar2022gptq,
  title     = {{GPTQ}: Accurate Post-Training Quantization for Generative Pre-trained Transformers},
  author    = {Frantar, Elias and Ashkboos, Saleh and Hoefler, Torsten and Alistarh, Dan},
  booktitle = {International Conference on Learning Representations (ICLR)},
  year      = {2023}
}

@article{dettmers2022llmint8,
  title   = {{LLM.int8()}: 8-bit matrix multiplication for transformers at scale},
  author  = {Dettmers, Tim and Lewis, Mike and Belkada, Younes and Zettlemoyer, Luke},
  journal = {Advances in Neural Information Processing Systems (NeurIPS)},
  year    = {2022}
}

@inproceedings{yao2022zeroquant,
  title     = {{ZeroQuant}: Efficient and Affordable Post-Training Quantization for Large-Scale Transformers},
  author    = {Yao, Zhewei and Yazdani Aminabadi, Reza and Zhang, Minjia and Wu, Xiaoxia and
               Li, Conglong and He, Yuxiong},
  booktitle = {Advances in Neural Information Processing Systems (NeurIPS)},
  year      = {2022}
}

@inproceedings{lin2024awq,
  title     = {{AWQ}: Activation-aware Weight Quantization for {LLM} Compression and Acceleration},
  author    = {Lin, Ji and Tang, Jiaming and Tang, Haotian and Yang, Shang and
               Chen, Wei-Ming and Wang, Wei-Chen and Xiao, Guangxuan and Dang, Xingyu and
               Gan, Chuang and Han, Song},
  booktitle = {Machine Learning and Systems (MLSys)},
  year      = {2024}
}

@article{ma20241bit,
  title   = {The Era of 1-bit {LLM}s: All Large Language Models are in 1.58 Bits},
  author  = {Ma, Shuming and Wang, Hongyu and Ma, Lingxiao and Wang, Lei and Wang, Wenhui and
             Huang, Shaohan and Dong, Li and Wang, Ruiping and Xue, Jilong and Wei, Furu},
  journal = {arXiv preprint arXiv:2402.17764},
  year    = {2024}
}

@inproceedings{rastegari2016xnor,
  title     = {{XNOR-Net}: {I}mage{N}et classification using binary convolutional neural networks},
  author    = {Rastegari, Mohammad and Ordonez, Vicente and Redmon, Joseph and Farhadi, Ali},
  booktitle = {European Conference on Computer Vision (ECCV)},
  year      = {2016}
}

@article{hubara2016binarized,
  title   = {Binarized neural networks},
  author  = {Hubara, Itay and Courbariaux, Matthieu and Soudry, Daniel and El-Yaniv, Ran and
             Bengio, Yoshua},
  journal = {Advances in Neural Information Processing Systems (NeurIPS)},
  year    = {2016}
}

@article{lin2014coco,
  title   = {Microsoft {COCO}: {C}ommon Objects in Context},
  author  = {Lin, Tsung-Yi and Maire, Michael and Belongie, Serge and Hays, James and
             Perona, Pietro and Ramanan, Deva and Doll{\'a}r, Piotr and Zitnick, C. Lawrence},
  booktitle = {European Conference on Computer Vision (ECCV)},
  year    = {2014}
}

@article{maierhein2018rankings,
  title   = {Why rankings of biomedical image analysis competitions should be interpreted with care},
  author  = {Maier-Hein, Lena and Eisenmann, Matthias and Reinke, Annika and Onogur, Sinan and
             Stankovic, Marko and Scholz, Patrick and Arbel, Tal and Bogunovic, Hrvoje and
             Bradley, Andrew P and Carass, Aaron and others},
  journal = {Nature Communications},
  volume  = {9},
  pages   = {5217},
  year    = {2018}
}

@article{maierhein2022metrics,
  title   = {Metrics reloaded: recommendations for image analysis validation},
  author  = {Maier-Hein, Lena and Reinke, Annika and Godau, Patrick and Tizabi, Minu D and
             Buettner, Florian and Christodoulou, Evangelia and Glocker, Ben and Isensee, Fabian and
             Kleesiek, Jens and Kozubek, Michal and others},
  journal = {Nature Methods},
  volume  = {21},
  pages   = {195--212},
  year    = {2024}
}

@article{reinke2024pitfalls,
  title   = {Understanding metric-related pitfalls in image analysis validation},
  author  = {Reinke, Annika and Tizabi, Minu D. and Baumgartner, Michael and
             Eisenmann, Matthias and B{\"a}nsch, Doreen and Kavur, A. Emre and
             Sudre, Carole H. and Acion, Laura and Antonelli, Michela and Arbel, Tal and others},
  journal = {Nature Methods},
  volume  = {21},
  pages   = {182--194},
  year    = {2024}
}

@article{varoquaux2022evaluating,
  title   = {How to evaluate performance in medical imaging: sample size, tests, and error bars},
  author  = {Varoquaux, Ga{\"e}l and Colliot, Olivier},
  journal = {arXiv preprint arXiv:2210.14677},
  year    = {2022}
}

@article{islam2018clt,
  title   = {Sample size and its role in the {C}entral {L}imit {T}heorem},
  author  = {Islam, Mohammed Rashedul},
  journal = {Computational and Applied Mathematics Journal},
  volume  = {4},
  number  = {1},
  pages   = {1--7},
  year    = {2018}
}

@book{efron1994bootstrap,
  title     = {An Introduction to the Bootstrap},
  author    = {Efron, Bradley and Tibshirani, Robert J.},
  year      = {1994},
  publisher = {Chapman \& Hall/CRC}
}

@article{diciccio1996bootstrap,
  title   = {Bootstrap confidence intervals},
  author  = {DiCiccio, Thomas J. and Efron, Bradley},
  journal = {Statistical Science},
  volume  = {11},
  number  = {3},
  pages   = {189--228},
  year    = {1996}
}

@article{nosek2018preregistration,
  title   = {The preregistration revolution},
  author  = {Nosek, Brian A. and Ebersole, Charles R. and DeHaven, Alexander C. and
             Mellor, David T.},
  journal = {Proceedings of the National Academy of Sciences},
  volume  = {115},
  number  = {11},
  pages   = {2600--2606},
  year    = {2018}
}

@article{pineau2021reproducibility,
  title   = {Improving reproducibility in machine learning research (a report from the {NeurIPS} 2019 reproducibility program)},
  author  = {Pineau, Joelle and Vincent-Lamarre, Philippe and Sinha, Koustuv and Larivi{\`e}re, Vincent and
             Beygelzimer, Alina and d'Alch{\'e}-Buc, Florence and Fox, Emily and Larochelle, Hugo},
  journal = {Journal of Machine Learning Research},
  volume  = {22},
  pages   = {1--20},
  year    = {2021}
}


\appendix




\end{document}